# Efficient Subgraph Matching on Billion Node Graphs

Zhao Sun[*]
Fudan University
zixiaojindao@gmail.com

Hongzhi Wang[*]
Harbin Institute of Technology
wangzh@hit.edu.cn

Haixun Wang
Microsoft Research Asia
haixunw@microsoft.com

Bin Shao
Microsoft Research Asia
binshao@microsoft.com

Jianzhong Li
Harbin Institute of Technology
lijzh@hit.edu.cn

## ABSTRACT

The ability to handle large scale graph data is crucial to an increasing number of applications. Much work has been dedicated to supporting basic graph operations such as subgraph matching, reachability, regular expression matching, etc. In many cases, graph indices are employed to speed up query processing. Typically, most indices require either super-linear indexing time or super-linear indexing space. Unfortunately, for very large graphs, super-linear approaches are almost always infeasible. In this paper, we study the problem of subgraph matching on billion-node graphs. We present a novel algorithm that supports efficient subgraph matching for graphs deployed on a distributed memory store. Instead of relying on super-linear indices, we use efficient graph exploration and massive parallel computing for query processing. Our experimental results demonstrate the feasibility of performing subgraph matching on web-scale graph data.

## 1. INTRODUCTION

Graphs serve as important data structures in many applications, including social network, web, and bioinformatics applications. An increasing number of applications work with web-scale graphs. For example, currently Facebook has 800 millions of vertices and the average degree of each vertex is 130 [2]. In 2000, the web graph already had 2.1 billion vertices and 15 billion edges [21]. Today, major search engines are laying out new infrastructure to support a web graph of 1 trillion vertices. In genome sequencing, recent work [31] attempts to solve the genome assembly problem by constructing, simplifying, and traversing the de Brujin graph of the read sequence. Each vertex in the de Brujin graph represents a $k$-mer, and the entire graph can contain as many as $4^k$ vertices where $k$ is at least 20.

Although a lot of efforts have been devoted to efficient graph query processing, few existing methods are designed for very large graphs, for example, billion-node graphs. To understand the challenge, consider various kinds of indices that have been developed to support graph query processing. Typically, indexing graphs is more complex than indexing relational data. State-of-the-art approaches employ indices of super-linear space and/or super-linear construction time. For example, the R-Join approach [9] for subgraph matching is based on the 2-hop index. The time complexity of building such an index is $O(n^4)$, where $n$ is the number of vertices. It is obvious that in large graphs where the value of $n$ is on the scale of 1 billion ($10^9$), any super-linear approach will become unrealistic, let alone an algorithm of complexity $O(n^4)$.

### 1.1 Subgraph Matching

In this paper, we focus on subgraph matching to highlight the critical situation we mentioned above, namely, many existing approaches for graph query processing do not work for very large graphs or billion-node graphs.

The subgraph matching problem is defined as follows: For a data graph $G$ and a query graph $Q$, retrieve all subgraphs of $G$ that are isomorphic to $Q$. Subgraph matching is one of the most fundamental operators in many applications that handle graphs, including protein-protein interaction networks [15, 36], knowledge bases [19, 27], and program analysis [34, 33]. In the following, we categorize representative graph matching approaches [26, 11, 13, 15, 9, 32, 36, 34] into 4 groups, and summarize their performance in Table 1.

1. **No index.** Some early methods perform graph-matching-like tasks without using index. The corresponding algorithms [26, 11] have super-linear complexity, and they only work for "toy" graphs, that is, graphs whose sizes (number of vertices) are on the scale of 1K.

2. **Edge index.** To support SPARQL queries on RDF data, RDF-3X [23], BitMat [5] and many other approaches create index on distinct edges. A SPARQL query is disassembled into a set of edges, and final answers are produced by multi-way joins. The problems with this approach are i) the excessive use of costly join operations, and ii) SPARQL can only express a subset of subgraph queries.

3. **Frequent subgraph index.** To avoid excessive joins, another approach is to find frequent subgraphs, or frequently queried subgraphs, and index these frequent structures [28]. A recent work in this direction is SpiderMine [35]. The problems with these approaches are i) finding frequent subgraphs is very costly, ii) a large number of frequent subgraphs lead to large index size, and iii) queries that do not contain frequent subgraphs are not very well supported.

---
[*]This work was done while the authors visited Microsoft Research Asia.

| | Algorithms | Index Size | Index Time | Update Cost | Size of graphs in experiments | Index Size for Facebook | Index Time for Facebook | Query Time on Facebook (s) |
|---|---|---|---|---|---|---|---|---|
| 1 | Ullmann [26], VF2 [11] | - | - | - | 4,484 | - | - | >1000 |
| 2 | RDF-3X [23] | $O(m)$ | $O(m)$ | $O(d)$ | 33M | 1T | >20 days | >48 |
|   | BitMat [5] | $O(m)$ | $O(m)$ | $O(m)$ | 361M | 2.4T | >20 days | >269 |
| 3 | Subdue [16] | - | Exponential | $O(m)$ | 10K | - | > 67 years | - |
|   | SpiderMine [35] | - | Exponential | $O(m)$ | 40K | - | > 3 years | - |
| 4 | R-Join [9] | $O(nm^{1/2})$ | $O(n^4)$ | $O(n)$ | 1M | >175T | $> 10^{15}$ years | >200 |
|   | Distance-Join [36] | $O(nm^{1/2})$ | $O(n^4)$ | $O(n)$ | 387K | >175T | $> 10^{15}$ years | >4000 |
|   | GraphQL [15] | $O(m + nd^r)$ | $O(m + nd^r)$ | $O(d^r)$ | 320K | >13T($r$=2) | > 600 years | >2000 |
|   | Zhao [34] | $O(nd^r)$ | $O(nd^r)$ | $O(d^L)$ | 2M | >12T($r$=2) | > 600 years | >600 |
|   | GADDI [32] | $O(nd^L)$ | $O(nd^L)$ | $O(d^L)$ | 10K | $> 2 \times 10^5$T ($L$=4) | $> 4 \times 10^5$ years | >400 |
|   | **STwig (our approach)** | $O(n)$ | $O(n)$ | $O(1)$ | 1B | 6G | 33s | <20 |

Table 1: Existing Subgraph Matching Methods ($n$ and $m$ are the number of nodes and edges in a graph, and $d$ is the average degree of a node).

4. **Reachability/Neighborhood index.** Recent work [9, 36, 34, 15, 32, 9, 36, 7] indexes information such as global or local reachability in the graph. For example, R-Join [9] and Distance-Join [36] use 2-hop labeling schemes for their index structures. For a graph with $n$ vertices and $m$ edges, the optimal size of a 2-hop labeling scheme is $O(nm^{1/2})$ [9] and index time complexity is $O(n^4)$. For each vertex $v$, GraphQL [15] indexes the subgraph within radius $r$ of $v$. In the same spirit, Zhao et. al. [34] encode the labels of vertices within distance $r$ of $v$ into a signature, and then index the signature. The space requirement of such approaches is $O(nd^r)$, where $d$ is the average degree of each vertex. To support efficient filtering of the vertices based on the 2-hop scheme, the value of $r$ needs to be at least 2. The worst case space requirement is $O(n^2)$ and the time complexity of index construction is $O(nm)$ [34]. Another approach, known as GADDI [32], creates index for every two nodes within distance $L$ and satisfying a proposed inequality property. The value of the smallest $L$ suggested by the author is 4.

In Table 1, we compare the above approaches in terms of space and time complexity, the size of the data used in the original work, and their performance on the Facebook graph (800 million nodes and 100 billion edges[1]). First, we can see that except for RDF, most graphs used in their experimental studies are small. RDF approaches can handle relatively larger data because SPARQL queries have less expressive power than general subgraph matching, and they rely on excessive use of joins, which reduces runtime performance. Second, indexing has super-linear space and time complexity. Third, index update is also very costly. For R-Join [9] and Distance-Join [36], which are based on the 2-hop labeling scheme [7], the index update time is $O(n)$. For GraphQL [15] and Zhao et. al.'s approach [34], updating a vertex will result in the rebuilding of the signature index of all vertices with distance $r$ to the vertex. Finally, for web-scale graphs (Facebook), none of these approaches is feasible even if massive parallelism is employed.

We also show the performance of our approach (STwig) in Table 1. In our approach, no graph structure index is needed, and the only index we use is for mapping text labels to graph node IDs, and the index has linear size and linear construction time. Even for graphs as large as the Facebook, the index performance is superior. In the experimental study, we will also show the query performance. Typically, the response time of subgraph matching on billion node graphs is less than a few seconds.

---
[1] Here we assume the data is hosted in one machine and the computation is also carried out in one machine.

## 1.2 Challenges

It is important to understand the new challenges presented by billion-node graphs. A billion-node graph usually cannot reside in the memory of a single machine. This creates problems for efficient data access on graphs, since one distinguishing characteristics of graphs is that graph accesses have no locality: As we explore a graph, we invoke random, instead of sequential data accesses, no matter how the graph is stored. In other words, any non-trivial graph query will have poor performance if the locality issue is not properly addressed. For example, although relational models or key/value stores can be used to manage graph data, they do not provide efficient query support, because graph exploration (random accesses) is achieved through (multi-way) joins, and to support the joins, indices must be used. This becomes a performance bottleneck. Indeed, some considers a *native* graph database to be one that allows the access of a vertex's neighbors without the use of any (disk-based) index [1].

Thus, in order to support efficient online query processing, we provide techniques to address the locality issue and support efficient graph exploration. Current solutions fall short of this requirement. Several distributed and parallel data processing systems, including Pregel [22] and InfiniteGraph [3], have been proposed. However, they are designed for offline graph analytics, instead of online query processing. Other systems, including Neo4j [1] and HyperGraphDB [17], are not for web-scale graphs and have only primitive support for basic graph operations.

Moreover, locality is not the only issue for online query processing on billion-node graphs. Existing graph algorithms, including those listed in Table 1, assume that the graphs are stored in memory, which means locality is not an issue. However, even if a billion-node graph is memory resident, existing algorithms are still insufficient for supporting basic graph operators such as subgraph matching. The reason is that, as shown in Table 1, they rely on indices with time and space complexity for billion-node graphs beyond our reach. In light of this, to support web-scale graphs, we cannot assume we can rely on sophisticated, super-linear indices. Only light-weight indices (linear or sublinear) that are easy to construct or maintain are feasible.

Then, how do we support efficient graph query processing when a web-scale graph cannot fit in the memory of a single machine, and no graph indices, or only lightweight index, is available? To our knowledge, this is a challenge that has not been addressed in previous research. In this paper, we present a solution.

## 1.3 Our Contributions

We use efficient in-memory graph exploration, instead of expensive joins, to support subgraph matching on large graphs deployed on a memory cloud that comprises a cluster of commodity machines. To our knowledge, this is the first system that can perform online subgraph matching on billion-node graphs.

Unlike previous work that relies on sophisticated indices, we perform subgraph matching without using index (except a "string" index that maps text labels to vertex IDs). This ensures the scalability of our algorithm for billion-node graphs, which are not *indexable*, both in terms of index space and index time. To make up for the performance loss due to lack of indexing support, we use intelligent graph exploration to replace expensive join operations. In our work, the basic graph exploration mechanism is provided by the Trinity [24, 25] memory cloud. A large graph is divided into multiple partitions and each partition is stored in the memory of a commodity machine. Trinity provides a transparent interface so that users can operate on the entire graph as if it were in a single memory space.

Given a query, we split it into a set of subquery graphs that can be efficiently processed via in-memory graph exploration. We perform join operations only if they are not avoidable (when there is a cycle in the query graph). This dramatically reduces query processing cost, which is usually dominated by joins. Furthermore, we assure that our method produces no duplicate results in the cluster, which means that deduplication is unnecessary when merging results returned from different machines.

In summary, the contributions of this paper are the following:

1. We propose a subgraph matching method suitable for web scale graphs. In particular, our approach does not use graph index, and thus we do not need to worry about index space or index update cost.

2. We present query optimization strategies for query partitioning to avoid expensive join operations. In this paper, we assume we have no information about data statistics, although such statistics can be used directly to further improve the optimization strategy.

3. Our query processing mechanism consumes very little memory, which is important for computing in the memory cloud environment.

## 1.4 Paper Outline

The remaining part of this paper is organized as follows. We define the problem and describe the graph database system in Section 2. Section 3 discusses the strategies for subgraph matching. We present the overview of our method in Section 4. Section 5 proposes the query optimization for our method. We perform an experimental study in Section 6. Related work are summarized in Section 7. Section 8 draws the conclusions.

## 2. BACKGROUND

In this section, we discuss the background of the subgraph matching problem, and we introduce the Trinity infrastructure used to store very large graphs.

### 2.1 Subgraph Matching

We perform subgraph matching on a labeled graph. Let $G=(V, E, T)$ be a graph, where $V$ is the set of vertices, $E$ is the set of edges, and $T: V \rightarrow \Sigma^*$ is a labeling function that assigns a label to each vertex in $V$. Intuitively, the goal of subgraph pattern matching is to find all occurrences of a graph pattern in a large graph. More formally, we define the subgraph query, and the subgraph matching problem as follows.

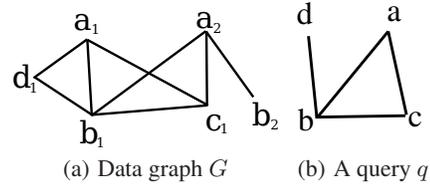

(a) Data graph $G$     (b) A query $q$

**Figure 1: The Example for Subgraph Matching**

DEFINITION 1 (SUBGRAPH QUERY). *We denote a subgraph query as $q=(V_q, E_q, T_q)$, where $T_q: V \rightarrow \Sigma^*$ represents the label constraint for each vertex in $V_q$.*

DEFINITION 2 (THE PROBLEM OF SUBGRAPH MATCHING). *For a graph $G$ and a subgraph query $q$, the goal of subgraph matching is to find every subgraph $g = (V_g, E_g)$ in $G$ such that there exists a bijection $f : V_q \rightarrow V_g$ that satisfies $\forall v \in V_q, T_q(v) = T_G(f(v))$ and $\forall e = (u,v) \in E_q, (f(u), f(v)) \in E_g$, where $T_G(f(v))$ represents the label of the vertex $f(v)$ in $G$.*

We use an example to illustrate subgraph matching. The data graph $G$ is shown in Figure 1(a), where each vertex has one label. In order to differentiate nodes with the same label, we append a suffix to the node. Thus, $a_i$ denotes the $i$-th node with label $a$. A subgraph query is shown in Figure 1(b). The results of the query are $(a_1, b_1, c_1, d_1)$ and $(a_2, b_1, c_1, d_1)$.

### 2.2 The Trinity Memory Cloud

As mentioned earlier, graph accesses have no locality. No matter how the graph is stored, we are likely to invoke random instead of sequential data accesses when we explore a graph. To ensure efficient random accesses, we can store the graph in RAM. However, very large graphs, such as the Facebook network and the Web, usually cannot fit in the RAM of a single machine.

We deploy graphs on Trinity [24]. Trinity is a memory cloud that comprises the RAM of one, dozens, or hundreds of machines. From users' perspective, Trinity provides a unified address space, as if a large graph is stored in the memory of one machine. Specifically, users explore the graph through APIs provided by Trinity. Issues related to storage, graph partitionings, and message passing are transparent to the user.

Memory cloud is consists of two core components: memory storage and network communication.

For storage, we use flat memory blob to store graph data instead of storing runtime objects on heap. Each runtime object on heap has non-neglectable meta data. In the extreme case, the meta data of a runtime object may even be larger than the payload data itself for small objects. The storage overhead of these meta data becomes very significant when many small objects exist in the system. For example, in one of our tests, 50 million 35-byte small objects takes 3.9 GB memory on CLR heap but only 1.6 GB in Trinity memory trunk. Concurrency control mechanisms are also carefully designed in Trinity. To boost system throughput, Trinity adopts a lock-free memory allocation mechanism [24]. Concurrent operations on a cell are coordinated using fine-grained cell spinning lock.

For network communication, Trinity adopts general network optimization principals such as message merging and batch transmission as well as techniques for minimizing message copy. On the sending side, message buffer and transmission buffer are reused when possible. On the receiving side, it is not easy to choose a proper buffer size since message sizes vary greatly. Trinity uses an extendable receiving buffer to minimize the overhead of buffer expansion and buffer shrinking.

Trinity provides very efficient support for graph exploration. In one experiment, we deployed a synthetic, power-law graph in a 15-machine cluster managed by Trinity. The graph has Facebook-like size and distribution (800 million nodes, 100 billion edges, with each node having on average 130 edges). We found that exploring the entire 3-hop neighborhood of any node in the graph takes less than 100 milliseconds on average. In other words, Trinity is able to explore $130 + 130^2 + 130^3 \approx 2.2$ million edges via RAM and network access in one tenth of a second. This lays the foundation for subgraph matching without using any structure index.

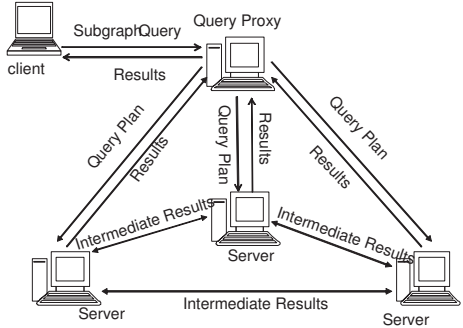

**Figure 2: Architecture Sketch**

The architecture of our subgraph matching system is sketched in Figure 2. First, a user submits a query. Then, we generate a query plan by decomposing the query. We send the query plan to the query proxy of the Trinity cloud. The proxy delivers the query plan to the Trinity slaves where the query plan is executed. The proxy coordinates the exchange of intermediate results, and sends back the results to the client after aggregating the results. Note that our algorithms ensure that the results sent back by different Trinity slaves are not overlapping, so the proxy does not need to perform deduplication.

## 3. EXPLORATION VS. JOINS

In this section, we discuss the alternative approach of using graph exploration, instead of substructure joins, to answer subgraph matching queries. The most significant benefits of graph exploration are: i) No structural index is needed; ii) Almost no join operation is required; and iii) We may reduce the size of intermediary results. These benefits are attractive for query processing on web scale graphs, where building and maintaining index is often an infeasible practice.

However, graph exploration is far from a carefree approach. In the following, we analyze the pros and cons of using join and graph exploration to answer subgraph matching queries.

### Using join operations for subgraph matching

The standard approach for subgraph matching usually proceeds as follows [4]. For a query graph $q$, we first decompose it into a set of smaller queries $q_1, \cdots, q_k$. Then we send the decomposed queries to the graph engine, which returns the results for each query. Finally, we join their results to answer the original query $q$. The rationale is that the decomposed queries can be answered by the graph engine directly. This is the case when the graph engine has an index entry that corresponds to each decomposed query. For example, assume we have indexed every unique edge in the graph. We will decompose a query into multiple edges.

However, the approach can be quite costly for the following two reasons: First, join operations are expensive. Second, a lot of intermediary results may be produced in vain. For example, the query in Figure 3(a) will be decomposed into two queries $(a, b)$ and $(b, c)$ and processed individually. Assume we perform this query on the graph in Figure 3(b). Using $b$ as the key, edges $(b_1, c_1)$ and $(b_1, c_2)$ are joined with $(a_1, b_1)$ to produce the correct results, but edges $(b_2, c_2), (b_3, c_2), \cdots, (b_k, c_2)$ are also generated and processed. It is obvious that if the query contains more edges, the problem will become more pronounced.

### Using graph exploration for subgraph matching

Now consider an alternative approach: We answer the query by graph exploration. Starting from query node $a$, we find $a_1$ through a simple index that maps labels to node IDs. We explore the graph to reach $b_1$, satisfying the partial query $(a, b)$. Then, from $b_1$ we explore the graph to reach $c_1$ and $c_2$, satisfying the partial query $(b, c)$. At this point, we have obtained the results without generating and joining large intermediary results. Certainly, if we start our exploration on nodes labeled $b$ or $c$, we might still get some useless intermediary results. But generally speaking, we will not generate as many of them unless it is in the worst case, and most important of all, join operations are avoided.

However, the graph exploration approach also has problems. First of all, it is not necessarily less costly. If the graph is stored in a relational database or a key/value store, then graph exploration itself requires join operations. In this paper, we perform graph exploration on the Trinity memory cloud to avoid this problem. Second, in some cases, naive graph exploration may be even more expensive than join operations. Consider the graph in Figure 3(c). Here, every $(a_i, b_j)$ and $(b_j, c_k)$ are part of the answer. The benefit of using join operations to answer this query is that with proper hash and merge strategies, the substructures can be joined in batch. With naive graph exploration, we need to traverse each path individually to produce the same set of results. Third, not all queries can be answered by graph exploration without using joins. To see this, consider the query in Figure 3(d). Assume we explore from node $a$ to $b$, then $c$ and $d$. At node $d$, we need to check if the next $a$ we see was the $a$ we started with. This is equivalent to a join operation. Furthermore, there are cases where both the join approach and the graph exploration approach perform badly. Consider performing query $q_2$ on graph $G_3$ in Figure 3(e). Here the only solution is $(a_1, b_1, c_1, d_m)$. However, the join operation will produce a lot of useless intermediary results, and the graph exploration approach will traverse many unfruitful paths.

### Weighting the pros and cons

Although the naive graph exploration method has many issues, its potential benefit – answering subgraph queries without using structure index – is attractive for query processing on web scale data. In this paper, we develop a novel graph exploration method that maximize the benefits of both the join approach and the naive graph exploration approach and avoid their disadvantages. Specifically, our method uses the join approach as the skeleton for query plan, and uses the exploration approach to avoid useless candidates during the join.

## 4. FRAMEWORK OF OUR APPROACH

In this section, we propose the framework for subgraph matching on billion-node graphs without using indexing the graph structure. For a subgraph query, we decompose it into a set of basic query units called *STwig*s. First, we show that *STwig* queries can be answered very efficiently in the memory cloud. Then, we describe the main steps of answering subgraph queries through *STwig*-based graph exploration and joining. Finally, we show how these steps are carried out in a distributed, parallel graph computing environment based on the Trinity memory cloud.

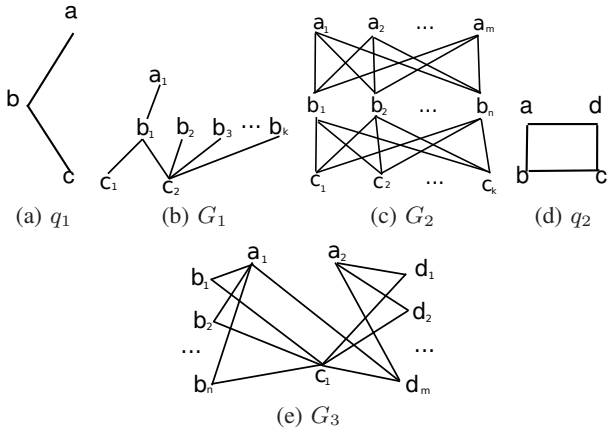

Figure 3: Examples for Discussions

### 4.1 Basic unit of graph access: STwig

With graphs stored in the memory cloud, and a string index mapping node labels to node IDs, the system provides the following atomic graph operators:

- `Cloud.Load(`$id$`)` : Locate the node whose ID is $id$. It also returns the IDs of the neighbors of the node.

- `Index.getID(`$label$`)`: Return the IDs of nodes with a given $label$.

- `Index.hasLabel(`$id$`, `$label$`)`: Return TRUE if the node whose ID is $id$ has a given $label$.

With the above basic operators, we implement a `MatchSTwig()` function. Specifically, an STwig is a two level tree structure. We use $q=(r, L)$ to denote an STwig query, where $r$ is the label of the root node and $L$ is the set of labels of its child nodes (which means the tree has only two levels). For example, in Figure 4(b), we have $q_1 = (a, \{b, c\})$, where "a", "b", "c" are labels.

Given $q$, Algorithm 1 finds matching STwigs in the graph in three steps: (1) Find the set of root nodes by calling `Index.getID(r)`; (2) For each root node, find its child nodes using `Cloud.Load()`; and (3) Find its child nodes that match the labels in $L$ by calling `Index.hasLabel()`. The function returns a set of STwigs that match the query. As an example, if we query $q_1$ against the data graph in Figure 5, the results are:

$$G(q_1) = \{ \;\; (a_1, b_1, c_1), (a_1, b_4, c_1), (a_2, b_1, c_1), (a_2, b_1, c_2),$$
$$(a_2, b_1, c_3), (a_2, b_2, c_1), (a_2, b_2, c_2), (a_2, b_2, c_3),$$
$$(a_3, b_2, c_2), (a_3, b_2, c_3)\}$$

The above function can be implemented very efficiently in the memory cloud. Now, given a subgraph query, if we decompose it into a set of STwigs, and use `MatchSTwig` to find matches to each STwig, then we can join their results to find the final solution. The approach can be extremely costly because each STwig is a very small structure, thus, each `MatchSTwig` may generate a large number of results. This also leads to a large number of joins, and

**Algorithm 1** MatchSTwig($q$) where $q = (r, L)$

$S_r \leftarrow$ `Index.getID`($r$)
$R \leftarrow \varnothing$
**for** each $n$ in $S_r$ **do**
  $c \leftarrow$ `Cloud.Load`($n$)
  **for** each $l_i$ in $L$ **do**
    $S_{l_i} \leftarrow \{m | m \in$ c.children and `Index.hasLabel`$(m, l)\}$
  $R = R \cup \{\{n\} \times S_{l_1} \times S_{l_2} \times \cdots \times S_{l_k}\}$
Return $R$

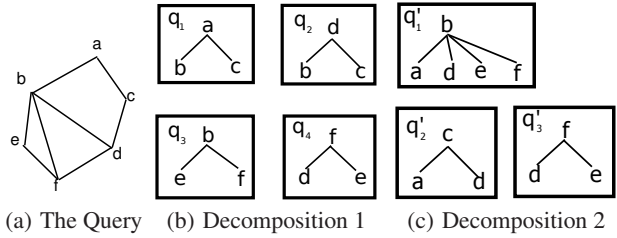

(a) The Query  (b) Decomposition 1  (c) Decomposition 2

Figure 4: A query and its 2 possible decompositions

very large amount of intermediary results. Traditional approaches try to break the original query into larger query units, which means we need to have structure index for these query units. As we mentioned, we cannot afford structure index for billion-node graphs. On the other hand, since our graph is memory-based, graph exploration is very fast. Thus, we implement a graph exploration based solution by searching in the neighborhood of matched STwigs to avoid the cost of joining many large set of results.

### 4.2 Subgraph Matching

Using STwig as the unit of query processing, our method performs subgraph matching in three steps.

#### 1. Query Decomposition and STwig Ordering

We decompose a query into a set of STwigs. We use an example to illustrate query partition. To process the query in Figure 4(a), we decompose it into the STwigs as shown in Figure 4(b). Another possible decomposition is shown in Figure 4(c), which contains only 3 STwigs. Clearly, different decompositions will incur different query processing cost. We discuss the decomposition strategy in Section 5.1.

Instead of finding matches for each STwig independently and join their results, we use graph exploration to perform the query. Thus, we create a linear order of the STwigs: $q_1, q_2, \cdots q_k$. The exploration will be conducted in this order. Clearly, the ordering has impact on the cost of query processing. We discuss the ordering strategy in Section 5.2 and 5.3.

#### 2. Exploration

We process an ordered list of STwigs one by one. Take the query in Figure 4(a) against the data graph in Figure 5 as example. Assume the query has been decomposed into the set of STwigs shown in Figure 4(b), and assume $q_1$ is the first STwig in the order produced in the previous step.

We process $q_1$ first, and we get result $G(q_1)$ in Eq 1. Next, we extract binding information from $G(q_1)$. For presentation simplicity, we assume nodes in the query graph are uniquely labeled. This allows us to use a label to address a unique query node. Thus, we do not need to involve node IDs in the following discussion.

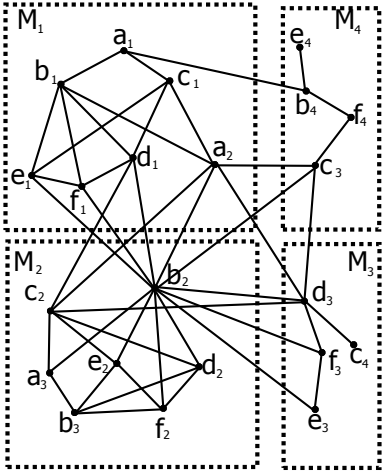

**Figure 5: A data graph partitioned over 4 machines:** $M_1, \cdots, M_4$.

For node (labeled) $a$ in $q_1$, let $H_a$ denote the binding information for $a$, that is, the set of nodes that match $a$. We have $H_a = \{a_1, a_2, a_3\}$. Similarly, we have $H_b = \{b_1, b_2, b_3, b_4\}$, $H_c = \{c_1, c_2, c_3\}$. The other nodes in query Figure 4(a) are not bound. That is, $H_d$ contains the set of all nodes in the data graph that match $d$. The binding information is only used to tell which nodes are eligible for the next queries. They cannot produce answers on their own, that is, $G(q_1) \neq H_a \times H_b \times H_c$.

Then, we process the next query in the predefined order. There are two cases. In the first case, suppose the next query in the predefined order is $q_2$. The root of $q_2$, which is labeled $d$, is not bound by the previous query $q_1$. So we need to invoke `Index.getID(d)` to find all nodes labeled $d$. However, the child nodes of these nodes are bound by $H_b$ and $H_c$. Thus, only nodes in $H_b$ and $H_c$ are eligible as their child nodes. Let $G(q_2)$ be the result of $q_2$. We update the binding information by incorporating the new bindings introduced by $q_2$.

In the second case, suppose the next query after $q_1$ is $q_3$. Here, the root of $q_3$ is already bound by $H_b$. Thus, instead of invoking `Index.getID(b)`, we use nodes in $H_b$ as the root nodes, and from there, we find their child nodes labeled $e$ and $f$. Let $G(q_3)$ be the result of $q_3$. We update the binding information accordingly. It is clear that the order of STwigs has impact on query efficiency. In particular, for a STwig $q$, if its root is bound by previous STwigs (the 2nd case above), fewer nodes need to be loaded. We will discuss STwig ordering in Section 5.2.

We process all STwigs in this manner until all of the nodes in the query are bound. The process generates a sequence of results $G(q_1), G(q_2), \cdots, G(q_k)$, which will be further processed (joined) in step 3.

*3. Join*

The previous step produces $G(q_1), \cdots, G(q_k)$, a sequence of intermediary results, one for each STwig. In this step, we join these results to produce the final answer. Clearly, each $G(q_i)$ is already much smaller than `MatchSTwig`$(q_i)$, since a lot of filtering and pruning is performed during exploration.

In this step, we perform two optimizations: join order selection and pipeline join. One question is why not join $G(q_1), \cdots, G(q_k)$ one by one as we produce them in the previous step, but instead, we postpone the join to the next step. The answer is that if we join the results in the predetermined order, it may generate unnecessarily large intermediate results. If the join is performed after all results of STwigs are generated (In a distributed environment, each machine also needs to fetch intermediary results produced by other machines. See Section 4.3 for a detailed discussion.), we can select a join order that minimizes the intermediary results. We apply sample-based join cost estimation method and cost-based join order selection method [14] to select the join order on each computer.

Even after all of the above optimizations, for certain queries, the size of intermediary results may still be quite big, which affects the online response time. This poses a significant challenge because our system is memory-based. To address this problem, we perform "block-based" pipelined join so that partial results are produced before the entire multi-way join is completed. Specifically, we divide the join into multiple rounds and in each round for each STwig, only a subset of its results (a block) participates the join. We use available memory to control the block size.

### 4.3 Distributed, Parallel Subgraph Matching

Our approach can be easily parallelized. The data graph is partitioned across multiple machines. In our work, we do not rely on any particular graph partitioning mechanism. In fact, our performance results are obtained in the setting where the graph is randomly partitioned (each node in the data graph is assigned to a machine by a hashing function). The string index in each machine only maps node labels to IDs of local nodes.

To parallel subgraph matching, we need to parallelize the 3 steps we outlined above. It is easy to handle the the first and second steps.

1. *Query Decomposition and STwig Ordering.* We do not parallelize this step. The proxy server decomposes the query graph and orders the STwigs, and then it broadcasts the results to all of the machines in the cluster.

2. *Exploration.* Each machine performs Algorithm 1 for STwig matching in parallel. Note that `Index.getID(r)` as well as `Cloud.Load(r)` only load data in the local machine. However, when checking the label of a child node $m$ by invoking `Index.hasLabel(m, l)`, we may incur network communication, because the child node may reside on a different machine. The memory cloud handles the memory and network access for these functions.

The parallelization of the 3rd step (join) is not trivial, and it is the focus of our discussion. After STwig matching, each machine $k$ produces $G_k(q_1), \cdots, G_k(q_n)$ for STwigs $q_1, \cdots, q_n$. We cannot just join them by themselves, instead, they need to join with results produced on other machines. Assume for STwig $q_i$, machine $k$ obtains multiple results from other machines and union them together:

$$R_k(q_i) = \bigcup_{k' \in F_{k,i} \cup \{k\}} G_{k'}(q_i)$$

where $F_{k,i}$ is the set of remote machines that machine $k$ need to access for their matches of $q_i$. We call $F_{k,i}$ the *load set*. Then, machine $k$ performs join on the obtained partial results:

$$R_k = R_k(q_{i_1}) \bowtie R_k(q_{i_2}) \bowtie \cdots \bowtie R_k(q_{i_n})$$

where $q_{i_1}, \cdots, q_{i_n}$ is a reordering of $q_1, \cdots, q_n$ determined by join order selection based on the statistics of the partial results. Finally, the results from different machines are unioned together to produce the answer to the subgraph matching query:

$$\bigcup_{k \in \text{cluster}} R_k \quad (1)$$

The question is, how do we decide *load set* $F_{k,i}$, that is, what other machines should machine $k$ access for their matches of STwig $q_i$? The choice of $F_{k,i}$ has two criteria. First, it is preferable that $R_k$'s are disjointness, that is, $R_k \cap R'_k = \varnothing$ for $k \neq k'$ so that when we compute the union we do not need to perform deduplication. Second, it is preferable that each $F_{k,i}$ is as small as possible, so that network communication is reduced.

The first criterion (disjointness) is easy to satisfy. We can pick a particular STwig (say $q_h$), which we call the *head* STwig, and set $F_{k,h} = \varnothing$ for every machine $k$. In Section 5.3, we show that the choice of *head* STwig has impact on query performance, and we will discuss in detail how to choose the most preferable *head* STwig. Here, we focus on showing how to ensure disjointness in answers. Since $F_{k,h} = \varnothing$, we have $R_k(q_h) = G_k(q_h)$, that is, the matching results on each machine for STwig $q_h$, which are to be joined with matching results of other STwigs, are the local results. Since the data graph is disjointly partitioned among the machines in the cluster, we know the local results are disjoint, that is, $G_k(q_h) \cap G'_k(q_h) = \varnothing$ for $k \neq k'$. This guarantees that $R_k$'s are disjoint, since for two answers that come from two different machines, the subgraphs in the answers that match $q_h$ are different (in this work, homomorphic answers are considered as different answers).

The next question is, for STwig $q_i$ other than the *head* STwig, what should *load set* $F_{k,i}$ be? To answer this question, consider a more specific question: Is there a simple way to find out that the matches of $q_i$ in machine $k$ will never join with matches of $q_{i'}$ in machine $k'$? Although the matches of a STwig may be all different, their root nodes have the same label. If a constraint between the root node of STwig $q_i$ and the root node of STwig $q_{i'}$ in the query is not satisfied, then we know for sure that their matches will not join. This is demonstrated by the following example.

EXAMPLE 1. *Consider the STwigs in Figure 4(c) and the data graph in Figure 5. Will matches of STwig $q'_1$ in $M_1$ join with matches of STwig $q'_3$ in $M_3$? The answer is no. This is because all matches of STwig $q'_1$ have a root node labeled b, and all matches of STwig $q'_3$ have a root node labeled f. In the query graph Figure 4(a), there is an edge between b and f. Thus, in order for their matches in $M_1$ and $M_3$ to join, there must be some edges between b nodes in $M_1$ and f nodes in $M_3$. However, no such edges exist.*

Clearly, knowing the matches of $q_i$ in machine $k$ will never join with matches of $q_{i'}$ in machine $k'$ will help us determine $F_{k,i}$. We discuss in detail how *load sets* are decided in Section 5.3.

## 5. QUERY OPTIMIZATION

As mentioned in Section 4, we perform query optimization in 3 aspects: i) query decomposition, ii) order selection for STwig matching, and iii) head STwig and load set selection.

### 5.1 Query Decomposition

It is clear that the number of STwigs determines the number of join operations. When performing the join operation, each machine needs to obtain intermediary results from other machines, which means processing more STwigs will incur more communication cost. On the other hand, loading an STwig with root node $r$ by Cloud.Load(r) has insignificant cost whether the STwig is big or small. Hence, the objective function of query decomposition is to minimize the number of components, that is, STwigs.

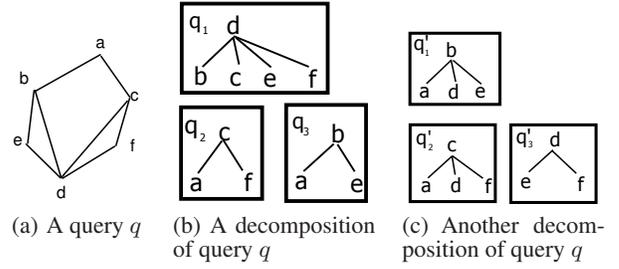

(a) A query $q$    (b) A decomposition of query $q$    (c) Another decomposition of query $q$

**Figure 6: STwig Order Selection**

PROBLEM 1. *Let $G$ be a query graph. Let $S=\{T_1,\cdots,T_n\}$ be a set of STwigs such that any edge of $G$ belongs to one and only one $T_i \in S$. We call $S$ an STwig cover of $G$. The problem is to find the minimum STwig cover of $G$.*

The following Theorem shows the hardness of finding the minimum STwig cover.

THEOREM 1. *The minimum STwig cover problem is polynomial equivalent to the minimum vertex cover problem.*

PROOF. Let $S$ be a minimum STwig cover of $G$. Let $V$ be the set of the root nodes of the STwigs. It is clear that $|V| = |S|$ and $V$ is a vertex cover of $G$, that is, any edge of $G$ is incident to at least one node in $V$. Next, we show $V$ is a minimum vertex cover. Assume there exists another vertex cover $V'$ such that $|V'| < |V|$. We construct an STwig cover $S'$ from $V'$ as follows. For each vertex $v \in V'$, we form an STwig which consists of $v$ and all edges incident to $v$. Then, we randomly delete edges in the STwigs until each edge belongs to one STwig. Let $S'$ be the set of STwigs that have at least one edge. Clearly, $|S'| \leqslant |V'| < |V| = |S|$, which means $S$ is not a minimum STwig cover. □

It is well-known that the minimum vertex cover problem is NP-hard. According to Theorem 1, the minimum STwig cover problem is also NP-hard. In the proof of Theorem 1, we showed that we can construct an STwig cover from a vertex cover in polynomial steps. There exists an 2-approximate algorithm [12] for the vertex cover problem. The algorithm works as follows. In each step, it randomly chooses an edge $(u, v)$, adds $u$ and $v$ in the result, and remove all edges incident to $u$ or $v$. It repeats the process until all of the edges are removed. We can use the same mechanism to create a 2-approximate STwig cover.

In our approach, we also need to find an optimal STwig processing order (Section 5.2). We will revise the query decomposition algorithm to ensure that the STwigs generated by the algorithm can lead to a more favorable STwigs processing order.

### 5.2 STwig Order Selection

Recall that the order of STwig processing may have significant impact on query efficiency (Section 4.2). Let $q$ be be the subgraph query as shown in Figure 6(a), and let Figure 6(b) be a possible decomposition of $q$. Now, consider the following two execution orders: $\langle q_1, q_2, q_3 \rangle$, and $\langle q_2, q_1, q_3 \rangle$. As we have made clear in our previous discussion, $\langle q_1, q_2, q_3 \rangle$ is a better order than $\langle q_2, q_1, q_3 \rangle$, because in the first order, when $q_2$ and $q_3$ are being processed, their root nodes ($c$ and $b$) are bound by the results of $q_1$, while in the second order, when $q_1$ is being processed, its root node ($d$) is not bound by the results of $q_2$.

It seems that given a set of STwigs produced by the decomposition step, we can simply choose an order that ensures, as much as possible, the root node of an STwig is bound by the results of previously processed STwigs. This approach, however, will not always lead to the optimal solution. Consider the two possible decompositions as shown in Figure 6(b) and 6(c). The two decompositions are of the same size, which means they are equally good. However, no matter how we re-order the STwigs for query processing, there always exist some STwigs whose root nodes are not bound by previously processed STwigs. For example, in $\langle q_1', q_2', q_3' \rangle$ or $\langle q_1', q_3', q_2' \rangle$, node $c$ of $q_2'$ is not bound. Similarly, in $\langle q_2', q_1', q_3' \rangle$ or $\langle q_2', q_3', q_1' \rangle$, node $b$ of $q_1'$ is not bound. Thus, the re-ordering mechanism cannot find an optimal order. But, such order exists, for example, $\langle q_1, q_2, q_3 \rangle$, which comes out of the decomposition of Figure 6(b).

Thus, in order to obtain a good STwig processing order, we revise the algorithm of query decomposition to generate decompositions that are more likely to produce good processing orders. Specifically, the revised 2-approximate algorithm ensures, except for the first STwig, the root of each STwig is a leaf node of at least one of the processed STwigs.

In the original approximate algorithm described in Section 5.1, edges are selected randomly. We revise the algorithm by adding two rules to guide edge selection.

1. Select edges that already connect to previously selected edges.
2. Select edges incident to nodes with high selectivity.

The first rule ensures that the root nodes of the STwigs are bound by previous STwigs. The second rule favors generating STwigs of higher selectivity in order to reduce the size of intermediary join results. The selectivity of an STwig is determined by two factors: i) the more popular its root node label, the lower selectivity the STwig has; and ii) the more child nodes its root rode has, the higher selectivity the STwig has. Hence, for each node $v$ in the query graph, we use $f(v) = \frac{deg(v)}{freq(v.label)}$ (called the $f$-value of $v$) to rank $v$, where $freq(l)$ is the number of nodes in the data graph that has label $l$.

**Algorithm 2** STwig-Order-Selection($q$)
1: $S = \varnothing$
2: $\mathbb{T} = \varnothing$
3: **while** $q$ has more edges **do**
4:   **if** $S = \varnothing$ **then**
5:     pick an edge $(v, u)$ such that $f(u) + f(v)$ is the largest
6:   **else**
7:     pick an edge $(v, u)$ such that $v \in S$ and $f(u) + f(v)$ is the largest
8:   $T_v \leftarrow$ the STwig rooted at $v$
9:   add $T_v$ to $\mathbb{T}$
10:   $S \leftarrow S \cup$ neighbor($v$)
11:   remove edges in $T_v$ from $q$
12:   **if** $deg(u) > 0$ **then**
13:     $T_u \leftarrow$ the STwig rooted at $u$
14:     append $T_u$ to $\mathbb{T}$
15:     remove all edges in $T_u$ from $q$
16:     $S \leftarrow S \cup$ neighbor($u$)
17:   remove $u, v$ and all nodes with degree 0 from $S$
18: **return** $\mathbb{T}$

Algorithm 2 outlines the method that combines query decomposition and STwig order selection. We use an example to demonstrate the algorithm. Let $q$ be the query graph in Figure 6(a), and assume each label matches 10 vertices in the data graph. In the first step, the set $S$ is empty, we choose an edge $(d, c)$, as $f(d) = 0.4$ and $f(c) = 0.3$ are among the largest. Then, we generate STwig $T_1 = \{d, (b, c, e, f)\}$ and STwig $T_2 = \{c, (a, f)\}$. After that, we have $S = \{a, b, e\}$. Note that $f$ is excluded from $S$ because its degree is 0 after all edges in $T_1$ and $T_2$ are removed. Now, $b$ has the largest $f$-value ($f(b) = 0.2$) and its two neighbors have the same $f$-values. We select edge $(b, a)$, and generate STwig $T_3 = \{b, (a, f)\}$. After $T_3$ is removed, no edge is left and the algorithm halts.

Regarding the time complexity of the algorithm, we note that computing $f$-values and sorting nodes by $f$-values have O($n \log n$) cost, where $n$ is the number of nodes in the query. In each round, at least two nodes are removed from $q$, so the iteration needs O($n$) steps. Thus, the time complexity of Algorithm 2 is O($n^2 \log n$).

Similar to the proof of the approximate ratio bound of VC problem [12], we prove the approximate ratio bound of Algorithm 2.

THEOREM 2. *Algorithm 2 is a 2-approximate algorithm. That is, the size of* $\mathbb{T}$ *is at most twice of the optimal solution to Problem 1.*

PROOF. Let $\mathbb{T}^*$ be the optimal STwig cover. Let $E$ be the edge set selected by our algorithm. Since any two edges in $E$ share no vertices and the edges in one STwig must share one vertex, each STwig contains at most one edge in $E$. It means that $|E| \leqslant |\mathbb{T}^*|$. Since in Algorithm 2, for each edge $e$ in $E$, at most two STwigs are generated with the roots as the two vertices of $e$, respectively. It implies that $|\mathbb{T}| \leqslant 2|E| \leqslant 2|\mathbb{T}^*|$. □

## 5.3 Head STwig and Load Set Selection

As we discussed in Section 4.3, head STwig and load set have big impact on communication cost. As we can see from Example 1, the communication cost is determined by how the data is distributed in different machines. In this section, we first introduce the cluster graph that models the data distribution, then we introduce how to decide the load set and choose the head STwig.

*The Cluster Graph*

Given a query, we create a *cluster graph* to model the data distribution among different machines in the cluster with regard to the query.

Specifically, for a data graph $G$ and a query $q$, we first introduce $G_q$, which is the part of $G$ that is relevant to $q$. The graph $G_q$ is created by removing edges in $G$ that do not match any edge in $q$. Clearly, the results of $q$ on $G$ and $G_q$ are the same. We then create a *cluster graph* $C$. Each vertex in $C$ uniquely represents a machine in the cluster. An edge $i \rightsquigarrow j$ exists in $C$ iff there exists an edge $u \rightsquigarrow v$ in $G_q$ such that $u$ and $v$ reside in machine $i$ and $j$ respectively. For example, let $G$ be the graph in Figure 5, and let $q$ be the query in Figure 4(a). We show $G_q$ in Figure 7(a) and the cluster graph $C$ in Figure 7(b).

One question is how costly it is to create a query-specific cluster graph? There is no need to materialize $G_q$. In the preprocessing phase, for each pairs of machines, we record all possible pairs of node labels. That is, we associate a pair of labels $(A, B)$ to a pair of machines $(i, j)$ if there exists an edge $u \rightsquigarrow v$ such that $u$ and $v$ reside in machine $i$ and $j$ respectively, and $u$ and $v$ are labeled $A$ and $B$ respectively. For a given query $q$, the cluster graph is constructed by looking up the stored label pairs for each edge in $q$ instead of accessing the data graph. The process can be carried out efficiently.

We now introduce how to use the cluster graph for optimization. Let $D_C(i, j)$ denote the shortest distance between two nodes $i$ and $j$ in the cluster graph $C$, and let $D_q(u, v)$ denote the shortest distance between $u$ and $v$ in $G_q$. We have the following theorem.

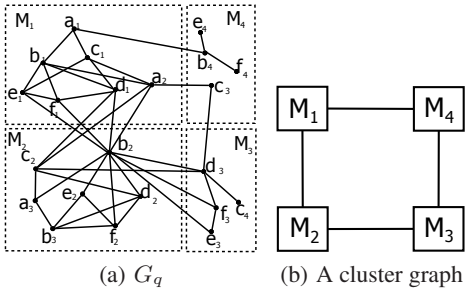

(a) $G_q$      (b) A cluster graph

**Figure 7: An Example for the Cluster Graph**

THEOREM 3. *For any nodes $u$ and $v$ reside in machines $i$ and $j$ respectively, we have $D_C(i,j) \leqslant D_q(u,v)$.*

PROOF. First, we show that the inequality holds when $D_q(u,v)=1$. There are two cases. In the first case, $u$ and $v$ reside in the same machine $i$. Obviously, $D_C(i,i) = 0 < D_q(u,v)$. Otherwise, there is an edge between machine $i$ and $j$ that match $u \rightsquigarrow v$. Since $D_q(u,v) = 1$, we have $D_C(i,j) = 1$. Next, we show that the inequality holds when $D_q(u,v) > 1$. Consider the path between $u$ and $v$ as a sequence of edges. For each edge $(u', v')$, the inequality is true. We then aggregate all the inequalities to obtain the inequality for the entire path. □

*Load Set*

Assume query $q$ is decomposed into a set of STwigs. For a machine $k$ and an STwig $q_t$ in the decomposition, we would like to find $F_{k,t}$, that is, the set of machines from which $k$ needs to request results of $q_t$. Assume $q_s$ is selected as the head STwig (the method of head STwig selection is given later in the section), which means each machine computes $q_s$ on its own without the need of communicating with other machines, we have the following result.

THEOREM 4. *Machine $k$ only needs query results for STwig $q_t$ from machines in the set $F_{k,t}$, which is given by:*

$$F_{k,t} = \{j | D_C(k,j) \leqslant d(r_s, r_t)\}$$

*where $d(r_s, r_t)$ denotes the shortest distance in $q$ between $r_s$ and $r_t$ (the root nodes of $q_s$ and $q_t$).*

PROOF. Let $r_i$ denote the root node of STwig $q_i$. Assume two subgraphs $g_s$ and $g_t$ match $q_s$ and $q_t$, respectively, and in $g_s$ and $g_t$, $n_s$ and $n_t$ match $r_s$ and $r_t$, respectively. If $g_s$ and $g_t$ belong to a match for $q$, then there is a path between $n_s$ and $n_t$ matching the shortest path between $q_s$ and $q_t$ in $q$. Since any path between $n_s$ and $n_t$ is no shorter than $D_q(n_s, n_t)$, we have $D_q(n_s, n_t) \leqslant d(r_s, r_t)$. If $g_s$ and $g_t$ are generated in machine $k$ and $j$ respectively, from Theorem 3, we have $D_C(k,j) \leqslant D_q(n_s, n_t) \leqslant d(r_s, r_t)$. Thus we have $F_{k,t} = \{j | D_C(k,j) \leqslant d(r_s, r_t)\}$. □

As an example, consider the query shown in Figure 4(b) against the graph in Figure 5. Assume $q_1$ is selected as the head STwig. In the query graph, the distance between $a$ and $b$ is 1. From Theorem 4, we have $F_{M_1,q_3} = \{M_2, M_4\}$, which shows that the results of $q_3$ are unnecessary to be loaded from $M_3$ to $M_1$.

*Head STwig Selection*

We model head STwig selection as an optimization problem. The goal is to minimize total communication among the machines.

THEOREM 5. *Assume a query $q$ is decomposed as $q_1, \cdots, q_n$ and $q_s$ is selected as the head STwig. The communication cost is*

$$T(s) = \sum_{k \in C} |\{j | D_C(k,j) \leqslant d(s)\}| \qquad (2)$$

*where $d(s) = \max_{1 \leqslant i \leqslant n} \{d(r_s, r_i)\}$.*

PROOF. Since a machine $k$ can request the results of multiple STwigs from a machine $j$ by a single communication, the total number of communications for $k$ is the maximal size of $F_{k,q_i}$ ($1 \leqslant i \leqslant n$). Thus the total communication times in the cluster $C$ is computed as:

$$\begin{aligned} T(s) &= \sum_{k \in C} \max_{1 \leqslant i \leqslant n} \{|F_{k,q_i}|\} \\ &= \sum_{k \in C} \max_{1 \leqslant i \leqslant n} \{|\{j | D_C(k,j) \leqslant d(r_s, r_t)\}|\} \end{aligned}$$

For any machine $k$, the larger $d(r_s, r_t)$, the larger $|\{j | D_C(k,j) \leqslant d(r_s, r_t)\}|$. Thus the maximal $d(r_s, r_t)$ implies the maximal $|\{j | D_C(k,j) \leqslant d(r_s, r_t)\}|$. Therefore,

$$T(s) = \sum_{k \in C} |\{j | D_C(k,j) \leqslant \max_{1 \leqslant i \leqslant n} \{d(r_s, r_i)\}\}|$$

□

Thus, we should choose $q_s$ where $s = \underset{x}{\mathrm{argmin}}\, T(x)$ as the head STwig. From Eq 2, to minimize $T(s)$, we should minimize $d(s)$. To solve this problem, we compute $d(i)$ for each $q_i$ and select the $q_s$ with minimal $d(s)$. We first compute the shortest path between each pair of vertices in a query $q$ using the Floyd Algorithm [12]. The result is a matrix $M$ where each entry $M_{u,v}$ is the length of the shortest path between $u$ and $v$ in $q$. For each pair of STwigs $q_i$ and $q_j$, $d(r_i, r_j)$ is $M_{r_i, r_j}$. Then for each $q_i$, we compute $d(i)$, which is the maximal value of $M_{r_i, r_j}$, that is, $d(i) \leftarrow \max_{1 \leqslant j \leqslant n} M_{r_i, r_j}$. Finally, we choose the STwig with the minimal $d(s)$ as the head STwig $q_s$.

The time complexity of all-pair shortest paths computation is $O(m^3)$. The time complexity of the following choosing steps are $O(m^2)$, $O(m^2)$ and $O(m)$, respectively. Thus the time complexity of the head STwig selection algorithm is $O(m^3)$ where $m$ is the number of vertices in the subgraph pattern $q$.

## 6. EXPERIMENTAL EVALUATION

To verify the performance of STwig approach, we performed experiments on both real data sets and synthetic data sets.

### 6.1 Experiment Environment

We perform experiments on two clusters: cluster 1 consists of 8 machines. Each one has 32 GB DDR3 RAM and two 2.53 GHz Intel Xeon E5540 CPU, each has 4 cores and 8 threads. The network adapter is Broadcom BCM5709C NetXtreme II GigE. Cluster 2 consists of 12 machines. Each machine has 48 GB DDR3 RAM and two 2.67 GHz Intel Xeon E5650 CPU, each has 6 cores and 12 threads. Dual network adapters are installed on cluster 2, one is 1 Gbps HP NC382i DP Multifunction Gigabit Server Adapter and the other is 40 Gbps Mellanox IPoIB Adapter. All experiments are implemented using C# and compiled using .NET Framework 4. The operating system is Windows Server 2008 R2 Enterprise with service pack 1.

We generate two kinds of queries for each graph. One query set is generated using DFS traversal from a randomly chosen node. The first $N$ nodes are kept as the query pattern. The other query set is generated by randomly adding $E$ edges among $N$ given nodes. A spanning tree is generated on the generated query to guarantee it is a connected graph. The nodes of a query are labelled from a given label collection once it is generated. The default values of $N$ and $E$ are 10 and 20 respectively.

For each data graph, we generate 100 queries and record its average execution time. When using the pipeline join strategy, the program terminates after 1024 matches have been found.

## 6.2 Experiments on Real Data

Cluster 1 is used for the experiments on real data. We performed several experiments on the following two data sets:

- **US Patents**[2] This directed graph represents the reference relations between US patents. This graph contains 3,774,768 nodes and 16,522,438 edges. We use the property class as the label collection. The graph has totally 418 labels.

- **WordNet**[3] This graph represents the relations between English words. Parts of speech are used as node labels. This graph has 82,670 nodes, 133,445 edges and 5 labels.

*DFS Queries.* Node counts of the queries vary from 3 to 10. The Experimental results are shown in Figure 8(a). On both data sets, the query time increases significantly from node count 7. While the execution times of queries with 10 nodes are similar to or even lower than those of queries with 9 nodes. The reasons are as follows: On one hand, more STwigs and joins will be processed when the number of query nodes gets larger. That's why the query time increases as query size gets larger. On the other hand, the labels of real data are relatively dense, such that the intermediate results to be joined are close to the combination of the nodes matching the labels when query size is small. For larger query, more STwig queries and join operations may be performed, however the intermediate results often get smaller under our exploration strategy, thus the overall processing cost is reduced.

*Random Queries.* To study the relationship between query size and execution time, we performed a set of experiments with different query sizes. The experimental results are shown in Figure 8(b), where the node count $N$ varies from 5 to 15, and the corresponding edge count is $2N$. The experimental results show that the query time is almost linear with the number of query nodes. It is because most randomly generated queries have relatively small result sizes. Hence, except for the first few STwigs, the costs of processing most STwigs are close. The join of intermediate results starts from small candidate sets with our join order selection strategy, thus the costs of most join operations are close. Larger the query, more join operations. Hence, the query time is nearly linear with the query size.

We have performed a set of experiments with edge count $E$ varies from 10 to 20 to study the impact of edge density as shown in Figure 8(c). According to the experimental results, the number of edges has no significant impact on the query time. Under our partition strategy, the average number of STwigs does not increase with the edge number. Thus the average numbers of STwig and join operation are very close. As discussed above, for randomly generated queries, the query time is nearly linear with the number of STwigs and joins. Therefore, the edge number in a query has small impact on the query time.

[2] http://vlado.fmf.uni-lj.si/pub/networks/data/patents/Patents.htm
[3] http://vlado.fmf.uni-lj.si/pub/networks/data/dic/Wordnet/Wordnet.zip

**Table 2: The Time of Graph Loading**

| Node number (M) | 1 | 4 | 16 | 64 | 256 | 1024 | 4096 |
|---|---|---|---|---|---|---|---|
| Load Time (s) | 2 | 4 | 9 | 36 | 66 | 266 | 689 |

*Speed-up.* The relation between query time and machine number are shown in Figure 9(a) and Figure 9(b). The same query is performed with machine number varies from 1 to 8. Three observations can be obtained from the experimental results. 1) As expected, the query time is significantly reduced as machine number increases. 2) The query time decreases sub-linearly with the machine number. The reason is that more network traffic and synchronization cost will be incurred with more machines. 3) The speedup ratio for DFS queries is greater than that for random queries. Each machine has relatively light work load when processing random queries due to smaller query result sizes, thus the speedup for random queries is not as significant as that for DFS queries.

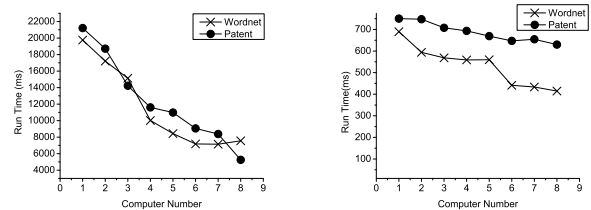

(a) Run Time VS. Computer Number (DFS)    (b) Run Time VS. Computer Number (Random)

**Figure 9: Speed-up on Real Data**

## 6.3 Experiments on Synthetic data

To verify the scalability of our approach, our algorithm is further evaluated on a set of synthetic graphs generated using R-MAT [8] model. The following experiments are performed on cluster 2. The default values of node count, average degree and average edge count are $64M$, 64 and 0.0001 respectively.

*Graph Size.* A set of experiments are performed to verify the scalability with regard to graph size. From the Table 2, even when the graph size scales to 1B, it can be loaded in our system within a few hundreds of seconds.

By the experimental results shown in Figure 10(a), graph size has no significant impact on the response time when average node degree is fixed (this degree is 16 in this experiment). Query response time varies between 400ms and 1800ms as graph node count varies from 1 million to 4 billion. No clear proportional relationship between response time and graph size is observed in this experiment. This indicates that the query time is not sensitive to total node count, thus this approach scales well as graph grows large. The reason is that the query time mainly depends on STwig number and STwig size, not the total number of the graph nodes. Experimental results shown in Figure 10(b) confirms this conclusion. In this experiment, node count varies with a fixed graph density. In this case, larger node count leads to larger average node degree. And larger

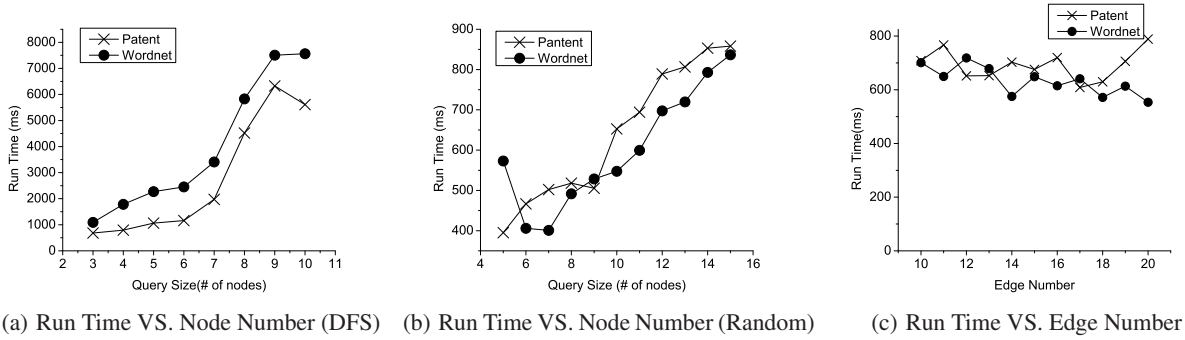

(a) Run Time VS. Node Number (DFS)  (b) Run Time VS. Node Number (Random)  (c) Run Time VS. Edge Number

Figure 8: Experimental Results on Real Data

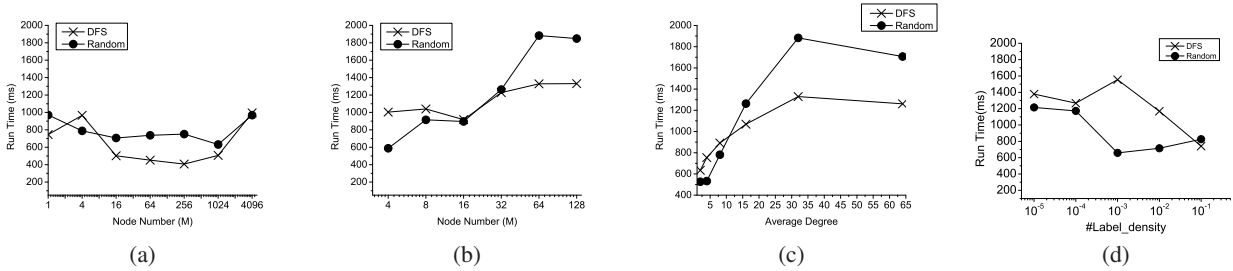

Figure 10: Experimental Results on Synthetic Data

node degree means larger STwig number and STwig size, thus the query time increases as the node count grows in this case.

*Graph Density.* The query response times under different graph densities are shown in Figure 10(c). We get two observations from the experimental results: 1) Greater the graph density, more neighbor nodes are accessed during STwig processing. No surprise, the query time increases as graph density grows large. However, the query time increases sub-linearly with graph density. 2) As graph gets dense, more intermediate results will be generated for randomly generated queries. Hence, graph density has greater impact on random queries than that on DFS queries as shown by the curve.

*Label Density.* Label density determines the number of the nodes that match the roots of STwigs. Higher label ratio, fewer matched nodes for a given label. Thus as expected, the query time decreases as label ratio varies from $10^{-5}$ to $10^{-1}$ as shown in Figure 10(d).

### 6.4 Summary

We summarize experimental results as follows.

1. Subgraph matching on billion node graphs can be completed within one or two seconds.

2. Our approach scales sub-linearly with graph size and graph density, thus it can scale to very large graphs.

3. Our approach can easily scale out. Larger graphs can be efficiently handled by adding more Trinity slaves.

### 7. RELATED WORK

Graph matching has been widely studied. We survey related work in two areas: graph pattern matching algorithms, and parallel and distributed graph processing techniques.

Subgraph pattern matching on graph data can be classified into queries on a large set of small graphs [28] and queries on a single large graph. The scalability issue in the first problem is mainly related to graph isomorphism, as the size of the graph set can be simply addressed by parallelism. The second problem is more challenging as far as the size of the graph is concerned. We study the second problem in this paper. Hence, this brief survey focuses on the queries on single large graph.

An important class of subgraph matching methods, called search-pruning methods [26, 11, 15, 30, 29], perform subgraph matching by searching the graph with some pruning strategies. For example, Ullmann [26] and Cordella et al [11] proposed global pruning strategy to accelerate query processing. However, the global pruning strategy requires the global information for the graph and not suitable for parallel processing for large graphs. Even though local pruning strategies are used in [15] to reduce the search space, the repetitive traversals of the graph cannot be avoided and cannot be applied to parallel processing.

Many approaches use index to accelerate subgraph pattern matching [9, 36, 32, 34, 18]. Cheng et al [9] and Zou et al [36] used 2-hop label scheme as index. GADDI [32] records the distance of each pair in the graph. The sizes of such indices are in more than square to the vertex number and the maintain cost is large. The indices in [33] and [34] use a signature representing neighborhoods within a given distance for each vertex. With a structural index costly to build and maintain, they cannot scale to web-scale data. Additionally, these index structures do not support parallelization.

It is natural to manage large graphs using parallel and distributed computation. Many parallel and distributed algorithms for graph problems, such as connected component, minimal spanning tree, all pairs shortest paths, etc. have been developed previously [20, 10, 6]. However, the study is mostly from an algorithmic perspective for parallel computing rather than from a big data perspective. They usually assume the graph can reside in the shared memory.

For example, traversal-based methods [10, 6] solve problems by parallelized graph traversal. On the other hand, matrix-partition-based methods [20] represent a graph as an adjacent matrix and parallelize the graph algorithms after partitioning the matrix onto multiple processors. Such kinds of methods require that the tasks can easily processed on the adjacent matrix, and they more suitable for dense graphs. Since subgraph matching is difficult to process on the adjacent matrix and most real-world large graphs are sparse, such methods cannot be applied for subgraph matching easily.

## 8. CONCLUSIONS

Subgraph matching on billion-node graphs is a challenge. One issue in previous approaches is that they all require index structures of super-linear size or super-linear construction time. This is infeasible for billion-node graphs. Our approach combines the benefits of subgraph join and graph exploration to support subgraph matching without using structure indices. For efficient query processing, we introduce several query optimization techniques, and a parallel version of our method. To deal with large intermediary results, we employ pipe-line join processing strategies. Experimental results demonstrate that our method can scale to billion-node graphs. We will do further experiments on even larger graphs (e.g. trillion scale graphs) to verify the system speedup, query throughput and response time bounds on massive data set. As another future work, some extra experiments will be also performed to test the amount of transmitted data on larger clusters.

**Acknowledgements.** Hongzhi Wang was supported by the Star Track program of Microsoft Research Asia. He and Jianzhong Li was partially supported by NGFR 973 grant 2012CB316200 and NSFC grant 61003046, 6111113089.